\documentclass{jfm}

\usepackage{graphicx}
\usepackage{newtxtext}
\usepackage{newtxmath}
\usepackage{natbib}
\usepackage[justification=justified,format=plain]{caption}
\usepackage{hyperref}
\hypersetup{
    colorlinks = true,
    urlcolor   = blue,
    citecolor  = black,
}

\newcommand{\RomanNumeralCaps}[1]
\linenumbers

\title{Undulatory underwater swimming: \\Linking vortex dynamics, thrust, and wake structure \\with a biorobotic fish}

\author{C. Brouzet\aff{1}
  \corresp{\email{christophe.brouzet@cnrs.fr}},
  C. Raufaste\aff{1,2}
 \and M. Argentina\aff{1}}

\affiliation{\aff{1}Universit\'e C\^ote d'Azur, CNRS, INPHYNI, France
\aff{2}Institut Universitaire de France (IUF), France}

\begin{document}
\maketitle

\begin{abstract}
Flapping-based propulsive systems rely on fluid-structure interactions to produce thrust. At intermediate and high Reynolds numbers, vortex formation and organization in the wake of such systems are crucial for the generation of a propulsive force. In this work, we experimentally investigate the wake produced by a tethered robotic fish immersed in a water tunnel. By systematically varying the amplitude and frequency of the fish tail as well as the free-stream speed, we are able to observe and characterize different vortex streets as a function of the Strouhal number. The produced wakes are three-dimensional and exhibit a classical V-shape, mainly with two oblique trains of vortex rings convecting outward. Using two-dimensional Particle Image Velocimetry (PIV) in the mid-span plane behind the fish and through extensive data processing of the velocity and vorticity fields, we demonstrate the strong couplings at place between vortex dynamics, thrust production and wake structure. We first measure the evolution of the vortex velocity with the Strouhal number, and model it using a momentum balance equation directly related to thrust production. We then focus on the wake structure, such as wake angle as well as vortex ring orientation, diameter and vorticity. The wake structure is modelled in a simple geometrical framework where the vortex ring velocity is composed of the free-stream speed and the ring self-advecting speed. This framework is tested and validated by our experimental measurements as well as literature data collapsing on master curves, highlighting a universal behavior dominated by the Strouhal number. This allows us to establish a comprehensive understanding of how the wake structure varies with this 
number and, thus, thrust production.
\end{abstract}



\section{Introduction}\label{sec:intro}

The wake of flapping objects has been widely studied in the last decades, from experimental, numerical or theoretical point of views. Indeed, when a flapping object oscillates in a free-stream flow, vortices are systematically shed and self-organise in vortex streets. The wake dynamics is known to be controlled by the dimensionless Strouhal number, defined as the ratio of the typical object tip oscillation speed to the free-stream speed.

A first step is to consider bi-dimensional (2D) wakes. These wakes are typically obtained in 2D environments, such as soap film experiments~\citep{SchnipperEtAl2009,AndersenEtAl2017} and 2D numerical simulations~\citep{DasEtAl2016,AndersenEtAl2017,ColvertEtAl2018}. They can also be studied in three-dimensional (3D) environments, for flapping objects with high aspect ratio between spanwise and streamwise dimensions~\citep{Koochesfahani1989,TriantafyllouEtAl1991,TriantafyllouEtAl1993,AndersonEtAl1998,Godoy-DianaEtAl2008,Godoy-DianaEtAl2009,BohlKoochesfahani2009,MackowskiWilliamson2015}. Most of the studies have considered the wake of a rigid oscillating foil, mainly pitching, but also heaving or with a pitching and heaving motion. The object is fixed at a given position in a free stream or has a steady forward motion in still water. The most observed vortical pattern closely resembles the typical B\'enard-von K\'arm\'an vortex street shed behind a bluff body (nicely illustrated by~\citet{VanDyke1982}). {At each oscillation period, the object sheds two single vortices of opposite vorticity, and this} pattern is therefore often named 2S in the literature (S~for single), following the nomenclature for the wake patterns behind a cylinder oscillating in a free stream~\citep{WilliamsonRoshko1988}. The organization of vortices within the 2S wake serves as a strong indicator of the force balance at play. With the convention of a flow originating from the left in a 2D plane, clockwise and counterclockwise vortices are shed above and below the centerline, respectively, in the drag regime (low Strouhal number), resembling the typical B\'enard-von K\'arm\'an wake~\citep{VanDyke1982}. As the Strouhal number increases and the relative contribution of thrust grows, both types of vortices join along the centerline and ultimately reverse positions above a given Strouhal number~\citep{Koochesfahani1989,Godoy-DianaEtAl2008,Godoy-DianaEtAl2009,BohlKoochesfahani2009}. Several works have shown that this vortex inversion in the wake occurs at a Strouhal number slightly smaller than the one where the transition between the {drag} and the {thrust} regimes happens~\citep{Godoy-DianaEtAl2008,Godoy-DianaEtAl2009,BohlKoochesfahani2009,DasEtAl2016,AndersenEtAl2017}. Regarding the time-averaged velocity fields, the vortex arrangement in the wake forms a jet behind the object~\citep{Koochesfahani1989}. The typical jet velocity is given by the positions of the clockwise and counterclockwise vortices with respect to the centerline: at low Strouhal number, below the vortex inversion, the jet velocity is smaller than the free-stream speed, while it becomes larger after the inversion, at large Strouhal number~\citep{TriantafyllouEtAl1991,TriantafyllouEtAl1993,Godoy-DianaEtAl2008,BohlKoochesfahani2009,AndersenEtAl2017}. The vortex inversion in the vorticity field therefore corresponds to a jet inversion in the time-averaged velocity field. Since the Strouhal numbers for the drag-thrust transition and the vortex inversion are very close, the time-averaged jet velocity also serves as a good indicator of the propulsion regime. In addition, note that other types of wakes have also been observed, mainly at smaller oscillation frequency~\citep{Koochesfahani1989,SchnipperEtAl2009,AndersenEtAl2017}. These wakes are more complex and are typically composed by two pairs of vortices shed per oscillation period, named 2P (P for pair), or by a combination of single vortices and vortex pairs, generally named as $m$S$+n$P, for $m$~single vortices and $n$~vortex pairs shed per oscillation period~\citep{WilliamsonRoshko1988}.

However, three dimensionality plays an important role in flapping systems when the aspect ratio of the moving object becomes small. In this configuration, the spanwise vortices, present in the 2D case, are now connected to vortices with an important streamwise vorticity component, thus leading to significantly more complex wakes with a 3D structure and organization~\citep{Eloy2012}. These wakes have been extensively studied for rigid flapping panels of low aspect ratio, both in experiments~\citep{vonEllenriederEtAl2003,ParkerEtAl2005,BuchholzSmits2005,BuchholzSmits2006,BuchholzSmits2008,BuchholzEtAl2008,BuchholzEtAl2011} and in numerical simulations~\citep{BlondeauxEtAl2005,DongEtAl2005,DongEtAl2006,LiDong2016}. These works have shown that there are three notably different wakes for low aspect ratio pitching panels: At low Strouhal numbers, the wake resembles a 3D transversely-growing B\'enard-von K\'arm\'an vortex street; With increasing Strouhal numbers, the wake bifurcates in a V-shape, forming two separate sets of vortex rings; At large Strouhal numbers, the wake keeps its V-shape but more complex vortical structures are shed and fill the entire wake region delimited by the bifurcation. For the two first types of wake patterns, the footprint in the mid-span plane of their 3D vorticity field resembles to the 2S and 2P vorticity fields observed for 2D flapping objects. By analogy, these 3D wakes are often called 2S and 2P in the literature. The 2S wake, extensively described and modelled by~\citet{BuchholzSmits2006,BuchholzSmits2008}, is composed of a succession of horseshoe vortices of alternated vorticity, interacting together with their streamwise leg. The 2P wake, the most observed wake type in the literature~\citep{vonEllenriederEtAl2003,BlondeauxEtAl2005,DongEtAl2005,DongEtAl2006,BuchholzSmits2008,LiDong2016}, has been discussed in detail and modelled by~\citet{DongEtAl2006} and \citet{BuchholzSmits2008}. It is composed of two oblique trains of vortex rings, which convect outward in the transverse direction. The orientation of the vortex rings with respect to the streamwise direction seems to indicate the propulsion regime (drag or thrust)~\citep{DongEtAl2006}. The 2P vortex rings originate from the closing of the 2S horseshoe vortices, due to the shedding of a spanwise shear layer from the trailing edge~\citep{DongEtAl2006,BuchholzSmits2008}. The transition between 2S and 2P wakes therefore appears smooth, and the wake angle increases uniformly with increasing Strouhal number~\citep{BuchholzSmits2006,LiDong2016}. Regarding the time-averaged velocity fields, as in the 2D geometry, the 2S and 2P wake patterns form jets behind the object, one streamwise jet for the 2S wake, and two oblique jets for the 2P wake. The mean velocity in the jets, depending on the vortex arrangement in the wake, seems to also indicate the regime (drag or thrust) of the propulsion~\citep{DongEtAl2006}. The topology of these two wakes appears to be relatively robust~\citep{DongEtAl2006,BuchholzSmits2008,LiDong2016} with respect to variation of the Reynolds number, of the shape of the object (rectangular or ellipsoidal) and of the aspect ratio of the object (within the low aspect ratio limit).

Undulatory underwater swimmers, such as fish and cetaceans, flap their tail to propel themselves through water. Despite the diversity in shape and size of their bodies, they definitely owns a finite-span tail and therefore naturally produce a 3D wake, like flapping foils of low aspect ratio. Fish wakes have been studied experimentally, behind real swimming fish~\citep{BlickhanEtAl1992,MullerEtAl1997,MullerEtAl2001,MullerEtAl2008,NauenLauder2002,TytellLauder2004,GuoEtAl2023} or robotic fish~\citep{ClarkSmits2006,EppsEtAl2009,GibouinEtAl2018}, and by several numerical simulations based on real fish body shapes and undulations~\citep{BorazjaniSotiropoulos2008,BorazjaniSotiropoulos2009,BorazjaniSotiropoulos2010,VanReesEtAl2013,BergmannIollo2016,LiEtAl2019a,LiEtAl2019,LiEtAl2022,GuoEtAl2023}. Given the undulating nature of fish's body and tail motion, the typical wakes observed are very similar to what has been described for oscillating panels of low aspect ratio. Indeed, both 2S and 2P wakes have been observed behind fish in the literature~\citep{TytellEtAl2010}: experiments performed with Particle Image Velocimetry (PIV) have mainly exhibited the footprint of the wakes in the mid-span plane, while numerical simulations have shown the full 3D structure of these two wakes. Slender-bodied fish, that tend to swim in the anguilliform mode where much of the body undulates at high amplitude, exhibit 2S wake structure. On the contrary, fish with broad tails, that tend to swim in the carangiform mode where the tail undulates at high amplitude, shed 2P wake pattern. However, \citet{TytellEtAl2010}, using the numerical works of~\citet{BorazjaniSotiropoulos2008,BorazjaniSotiropoulos2009,BorazjaniSotiropoulos2010}, have shown that both types of wake can be obtained on tethered fish, \textit{i.e.} when thrust and drag forces are not necessarily balanced. The transition between the two wakes strongly depends on the Strouhal number, while Reynolds number, swimming mode and body shape have smaller effects on the gross wake structure. Indeed, as for the rigid flapping foil of low aspect ratio, 2S wakes are obtained at low Strouhal number while 2P wakes are recovered at higher Strouhal number. Despite significant differences between the exact shape and motion of the flapping objects, the three-dimensional 2S and 2P wake patterns systematically appear for flapping foils of low aspect ratio and undulatory underwater swimmers. They therefore seems quite robust and universal. 

Thrust generation and propulsive motion are intimately related to vortex organisation within the wake~\citep{TriantafyllouEtAl1991}. Indeed, thrust production can be qualitatively determined~\citep{Godoy-DianaEtAl2008,DongEtAl2006,AndersenEtAl2017} by looking at a snapshot of the vorticity field, \textit{i.e.} vortex positions and vortex pair orientation, or by considering the time-averaged velocity field, \textit{i.e.} by comparing the streamwise component of the jet velocity with respect to the free-stream speed. Beyond these simple principles, thrust can be quantitatively evaluated through a momentum balance using the time-averaged velocity field~\citep{Batchelor1967,Koochesfahani1989,AndersonEtAl1998,Godoy-DianaEtAl2008,BohlKoochesfahani2009}. However, the links between thrust production, on one hand, and wake structure and vortex dynamics, on the other hand, have not been investigated in detail. This paper therefore proposes to fill this gap, by examining the different wakes produced by a tethered robotic fish using 2D PIV in the mid-span plane. By performing careful measurements of vortex and jet properties in that plane and by developing a geometrical model for the wake organisation, we establish a comprehensive understanding of the relationship between thrust production, vortex dynamics and wake structure.

This paper is organized as follows. Section~\ref{sec:setup} describes the experimental setup, the PIV system used in this study, the parameter space that has been explored and the different wakes observed. In Section~\ref{sec:characteristics}, we focus on the link between thrust production and vortex velocities, through measurements of the vortex and jet velocities as a function of the Strouhal number and a model using momentum balance. In Section~\ref{sec:V-shaped}, the organisation of vortex streets is discussed and modelled using a simple geometrical framework, allowing us to exhibit the link between vortex properties and wake structure. Conclusions are given in Section~\ref{sec:ccl}.
 
\section{Experimental setup and procedures}\label{sec:setup}

\subsection{Robotic fish platform}

\begin{figure}
  \centerline{\includegraphics[width=1\textwidth]{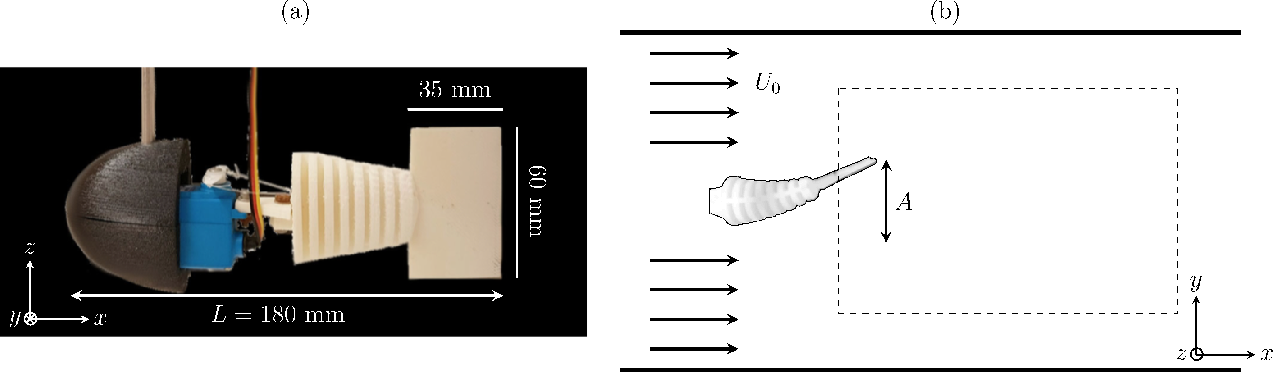}}
  \caption{(a) Robotic fish used in this study, seen in the streamwise ($x$) -  spanwise ($z$) plane. (b) Sketch of the configuration of the robotic fish in the water tunnel, seen from top in the streamwise ($x$) - transverse ($y$) plane. The imaging section of the PIV measurements is shown as a dashed rectangle of length~$180$~mm in the streamwise $x$-direction and width~$120$~mm in the transverse $y$-direction. The free-stream speed in the channel is~$U_0$, while the fish moves at frequency~$f$ and with a tail amplitude~$A$.}
\label{fig:setup}
\end{figure}

The experimental setup consists of a robotic fish placed at mid-depth in a laminar water tunnel (Rolling Hills Research Corporation, Model 0710). The robot, described in~\citet{GibouinEtAl2018}, \citet{Sanchez-RodriguezEtAl2021} and \citet{FuEtAl2023}, is shown in Fig.~\ref{fig:setup}(a). It is composed of a 3D printed hard head (on the left, in black), a servomotor (in the middle, in blue) and a 3D printed soft tail (at the right, in white). The fish has a total length of $L=180$~mm in the streamwise $x$-direction, a width of $27$~mm in the transverse $y$-direction, and a height of~$60$~mm in the spanwise $z$-direction. The aspect ratio of the tail, defined as the ratio between its respective sizes in the spanwise and streamwise directions, is approximately equal to $1.7$, which is in the same range than the objects with low aspect ratios studied in the literature~\citep{DongEtAl2006,BuchholzSmits2006,BuchholzSmits2008,LiDong2016}. The servomotor drives a wheel, where two cables are attached. Each cable goes on one side of the tail. When the servomotor wheel is rotating, the two cables alternatively pull on each side of the first part of the tail, thus mimicking antagonistic muscles and deforming it. A typical deformation of the tail, seen from the top, is visible in Fig.~\ref{fig:setup}(b). The servomotor wheel performs an oscillatory motion, leading to undulations of the fish at a frequency~$f$ and with a peak-to-peak amplitude~$A$ {for the tail}. The free-stream speed of the water tunnel~$U_0$ can be varied up to $110$~mm/s. The channel test section, with {a glass plate on the top for optical access}, has a length of $450$~mm, a height equal to $250$~mm and a width of $180$~mm. The end of the tunnel is also 
equipped with a transparent glass plate, allowing us to illuminate the test section behind the fish with a horizontal laser sheet. 

Measurements were performed using two-dimensional Particle Image Velocimetry (PIV) on the horizontal mid-span plane behind the robotic fish, with a camera filming from the top. The imaging section, shown as a dashed rectangle in Fig.~\ref{fig:setup}(b), measures $180 \times 120$~mm$^2$. {The origin of the coordinate system in the $x-y$ plane is taken at the tip of the fish tail, when the fish is immobile.} PIV acquisition and post-processing were done using a LaVision system (DaVis 10.2.0) with a Photron SA3 high-speed camera ($1024 \times 1024$~pixels) recording images at $200$~Hz and a Litron LDY 300 PIV pulsed laser. Laser sheet width was about $1$~mm in the whole imaging region. Additional post-processing and analysis were done using MATLAB and the PIVMat Toolbox~\citep{Moisy2007}.

The fish is hold fixed in the laboratory frame by its head, which is attached to a force sensor (visible as a gray rod on the left in Fig.~\ref{fig:setup}(a)). This allows us to measure the instantaneous net force~$F$ exerted on the fish, combining the drag force~$F_D$ of the fluid on the fish body and the thrust force~$F_T$ produced by the robot, such as $F=F_T-F_D$~\citep{GibouinEtAl2018}. This configuration, where the object is fixed in a given flow, is very classical to study the wakes produced by the object. It should be noted that mimicking the situation where a fish swims at a constant speed in a quiescent fluid is possible only when the time-averaged net force~$\langle F\rangle$ is equal to 0~\citep{BorazjaniSotiropoulos2010}. If the net force is negative (drag regime) or positive (thrust regime), this means that the fish would be decelerating or accelerating in the flow. However, exploring the different wakes produced as a function of the Strouhal number allows us to fully understand the effect of this parameter on the vortex street, thus being more general. 

\subsection{Parameter space}

\begin{table}
  \begin{center}
\def~{\hphantom{0}}
  \begin{tabular}{cccccc}
      Series  & ~$U_0$~[mm/s]   &   ~$A$~[mm]~& $f$~[Hz] & ~St$_A$ & Color code \\[3pt]
       1  & 103 & $13-55$ & $0.5-2$& ~$0.1-0.6$ & black\\
       2   & 53 & $10-44$ & $0.2-2$ & $0.1-1$ & red \\
       3   & 26 & $11-50$ & $0.5-2$ & ~$0.4-2.2$ & orange
         \end{tabular}
  \caption{Range of dimensional parameters and Strouhal numbers explored in this study.}
  \label{tab:param}
  \end{center}
\end{table}

\begin{figure}
  \centerline{\includegraphics[width=1\textwidth]{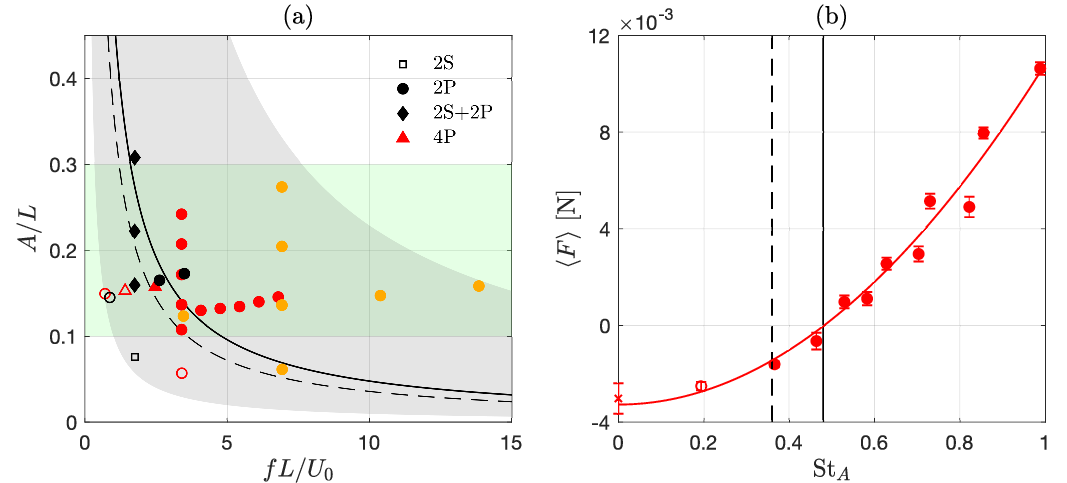}}
    \caption{(a) Kind of vortex streets observed in the parameter space~$(A/L,fL/U_0$). The color of each point corresponds to the free-stream speed~$U_0$ (orange~$U_0=26$~mm/s, red~$U_0=53$~mm/s, black~$U_0=103$~mm/s), while its shape represents the observed form of the wake, described in section~\ref{sec:obs}. The empty symbols indicate that the jets formed behind the fish have a time-averaged streamwise velocity component below the free-stream speed~$U_0$. The horizontal green band between $A/L=0.1$ and $0.3$ marks the limit observed for different fish species, while the gray region delimits the Strouhal number range explored in this study, between St$_A=0.1$ and~$2.2$. The solid curve represents the critical Strouhal number~St$_A^\star=0.48$, for which drag-thrust transition occurs for the robotic fish used in this study, and the dashed curve corresponds to the vortex or jet inversion at~St$_A^\textrm{inv}=0.36$. (b) Time-averaged force~$\langle F \rangle$ measured by the force sensor as a function of the Strouhal number~St$_A$, for 
  experiments of series~2 (see Table~\ref{tab:param}). A vertical solid line indicates the drag/thrust transition at St$_A^\star\approx 0.48$ and a vertical dashed line shows the vortex or jet inversion at ~St$_A^\textrm{inv}=0.36$. The red curve represents the fit given by Eq.~\eqref{eq:force_sensor}. The symbols are similar to the ones used in the other figures when the fish is moving, while the point at St$_A=0$, taken when the fish is immobile, is represented by a cross (x).}
\label{fig:param_space}
\end{figure}

While the length~$L$ of the fish is constant, a large parameter space has been explored by varying the three main quantities governing the dynamics of the wake: the fish tail amplitude~$A$ and frequency~$f$, and the free-stream speed~$U_0$. The explored ranges of the different parameters are summarised in Tab.~\ref{tab:param}. This parameter space can be quantified using several dimensionless quantities. The dimensionless amplitude~$A/L$ has been varied within the range $0.1-0.3$, which corresponds to what is observed on living fish and cetaceans~\citep{Bainbridge1958,HunterZweifel1971,RohrFish2004,SaadatEtAl2017,Sanchez-RodriguezEtAl2023}. The dimensionless frequency~$f L / U_0$ has been varied in a wide range, between $1$ and~$15$. As in the literature~\citep{Godoy-DianaEtAl2008,Godoy-DianaEtAl2009,SchnipperEtAl2009,AndersenEtAl2017,SaadatEtAl2017}, these two parameters are used to represent the different experiments on a parameter space map shown in Fig.~\ref{fig:param_space}(a). However, the main control parameter of the wake is the Strouhal number based on the fish tail amplitude, defined as St$_A\equiv f A /U_0$. This dimensionless parameter is the product of the two previously defined parameters, and combined in itself the three physical quantities varied in the experiments. Constant Strouhal curves in the diagram of Fig.~\ref{fig:param_space}(a) correspond to hyperboles. Here we have explored a large range of Strouhal number St$_A$, from $0.1$ to $2.2$, while for most of living fish and cetaceans, the Strouhal number is observed to be almost constant, between $0.2$ and $0.4$~\citep{TriantafyllouEtAl1991,GazzolaEtAl2014,SaadatEtAl2017,Sanchez-RodriguezEtAl2023}. We have chosen here to explore a much larger range to obtain the comprehensive scaling laws as a function of this control parameter, below and above the critical Strouhal number where the transition between the drag and thrust regimes occurs in our case~(St$_A^\star \approx 0.48$, see paragraph below). Note that the Reynolds number based on the robotic fish length, $\textrm{Re}=U_0 L /\nu$, where $\nu$ is the kinematic viscosity of the fluid, here water, is in between $4500$ and $18000$.

The time-averaged force~$\langle F \rangle$, measured by the force sensor, is shown in Fig.~\ref{fig:param_space}(b) as a function of the Strouhal number. Its trend is consistent with the curves shown in the literature~\citep{BohlKoochesfahani2009,TytellEtAl2010,MackowskiWilliamson2015}. 
At low Strouhal number, the fish is in the drag regime with $\langle F \rangle<0$. As the Strouhal number increases, the force also increases and becomes positive for a critical Strouhal number. This means that the averaged thrust~$\langle F_T \rangle= C_T \rho A^2 f^2 L^2$ produced by the fish is larger than the fluid drag $\langle F_D \rangle= \rho C_D U_0^2 L^2$. Here, $C_T$ is the a priori unknown proportionality constant between the real thrust and its scaling law~\citep{GazzolaEtAl2014,GrossEtAl2021} while $C_D$ is the drag coefficient of the fish~\citep{GibouinEtAl2018,Sanchez-RodriguezEtAl2021}. Note that the drag coefficient~$C_D$ is assumed to be the same when the robot is immobile and when it is moving, even though this coefficient is slightly higher when the fish is moving~\citep{BorazjaniSotiropoulos2010,GibouinEtAl2018,GrossEtAl2021}. At the critical Strouhal number~St$_A^\star$
, the averaged fish thrust balances the fluid drag and the force~$\langle F \rangle$ cancels. Therefore, the force~$\langle F \rangle$ can rewritten as
\begin{equation}
\langle F \rangle = \langle F_T \rangle-\langle F_D \rangle=\rho C_D L^2 U_0^2 \left( \left(\frac{\textrm{St}_A}{\textrm{St}_A^\star}\right)^2 - 1\right),\label{eq:force_sensor}
\end{equation}
with St$_A^\star=\sqrt{C_D/C_T}$. Note that this equation is fully compatible with the evolution of the time-averaged force~$\langle F \rangle$ with the Strouhal number, as shown by the fit of the experimental data in Fig.~\ref{fig:param_space}(b). This fit allows us to determine the value of critical Strouhal number for our experiments: St$_A^\star\approx0.48$.
This value is consistent with the critical Strouhal number reported for a very similar fish, build on the same principle, in a previous paper~\citep{GibouinEtAl2018}. Note that this value is higher than the average Strouhal number for which biological fish are found to swim (St$_A=0.2-0.4$~\citep{TriantafyllouEtAl1991}). This means that our robot is less efficient than real fish: indeed, its shape is not completely optimised regarding drag and its tail shape and motion do not reflect all the complexity found in nature. However, both values are of the same order of magnitude, and we here vary the Strouhal number in a large range around the critical Strouhal number.

\subsection{Observed wakes}\label{sec:obs}

\begin{figure}
  \centerline{\includegraphics[width=1\textwidth]{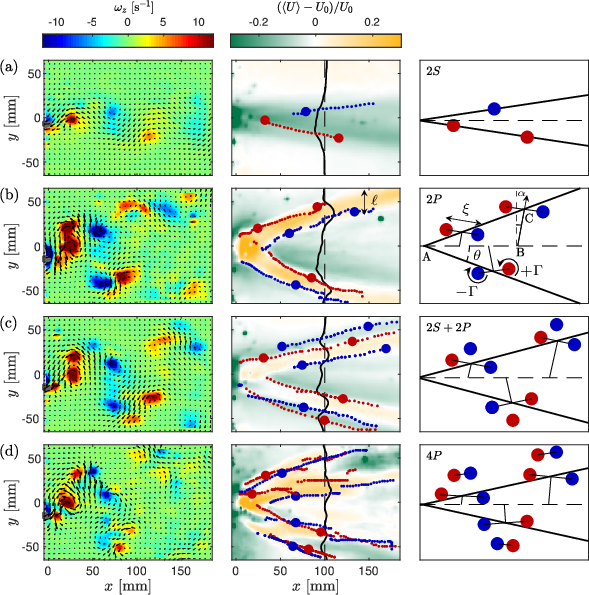}}
  \caption{Vorticity, velocity and wakes observed. Each row represents a typical wake: 2S (a), 2P (b), 2S+2P (c) and 4P (d). The left column shows the phase averaged vorticity (in color) and velocity (with arrows and where $U_0 \vec{e}_x$ is subtracted) fields when the tail is at its maximum amplitude. 
  The middle column represents the normalised time-averaged streamwise velocity field, obtained as $(\langle U \rangle-U_0)/U_0$, where $\langle U \rangle$ is the time-averaged streamwise velocity field. The trajectories of the vortices are represented on top of the normalised streamwise velocity, with red and blue dots. The color of the points represent clockwise (blue) or counter-clockwise (red) vortices. The position of the vortices visible in the left panels are marked by blue or red disks in the middle panels. 
  At $x=100$~mm, the time-averaged streamwise velocity profiles are plotted as black lines, illustrating deviations from the free-stream speed (dashed line), which indicate the presence of one or multiple jets. The right column shows sketches of the corresponding wakes, with the definitions of the notations given in panel (b): $\theta$ is the wake angle, $\alpha$ the orientation angle of the main vortex pair with the transverse $y$-direction, $\xi$ the pair diameter, i.e. the distance between the two vortices of the same pair, and $\pm\Gamma$ the circulation of each vortex.
}
\label{fig:4_wakes}
\end{figure}

Within the explored parameter space, four different types of wakes have been quantitatively observed, whose typical examples are given in Fig.~\ref{fig:4_wakes}. It is important to note that, since these wakes own an obviously 3D structure due to the low aspect ratio of the tail, we observe only the 2D footprint of these wakes in the mid-span plane behind the fish with the 2D PIV technique. Nevertheless, this is fully sufficient to classify the observed wakes in different categories. The first one corresponds to 2S wakes (Fig.~\ref{fig:4_wakes}(a)), already documented in 3D by \citet{BuchholzSmits2006,BuchholzSmits2008}. It is composed of 3D horseshoe vortices, whose footprint in the mid-plane corresponds to two~single vortices emitted per period. In the figures of the paper, the symbol representing this type of wake is a square. The second type of wake, the most commonly observed in our experiments, corresponds to 2P wakes (Fig.~\ref{fig:4_wakes}(b)), with 3D vortex rings emitted on each side of the fish~\citep{DongEtAl2006,BuchholzSmits2008}. Its signature in the mid-plane corresponds to two pairs of contra-rotating vortices emitted per period. This wake is represented by a circle in the following. Note that 2S and 2P wakes have already been observed for fish in several situations~\citep{TytellEtAl2010}, such as PIV experiments behind living fish or numerical simulations with a realistic fish body. The two other types of wakes are more complex, since they are composed by two~groups of three or four~vortices emitted per period in the mid-plane (Figs.~\ref{fig:4_wakes}(c) and (d)). Their 3D structure has not been described in the literature, but their 2D footprints in the mid-span plane correspond to a combination of the two first types of wakes. They are therefore named respectively 2P+2S and 4P, and represented respectively by a diamond and a triangle.

Despite their differences in structure and number of vortices, one can clearly see in Fig.~\ref{fig:4_wakes} that both vortex trajectories and time-averaged velocity field form a V-shaped or bifurcated wake behind the fish, whatever the type of wake. As already observed in the literature for 2S and 2P wakes, the time-averaged velocity field exhibits jets behind the fish. Here we have between one and four jets, depending on the wake type. For each wake, the time-averaged streamwise velocity 
within the jets can be smaller (represented by open symbols in Fig.~\ref{fig:param_space}) or larger (solid symbols in Fig.~\ref{fig:param_space}) than the free-stream speed~$U_0$. In all cases, these jets are delimited by the vortex trajectories, that typically mark the transition where the time-averaged streamwise velocity field 
is equal to $U_0$ (white lines with the colormap used). This clearly shows that the jets in the time-averaged velocity field are created by the  periodic passing of the vortex structures, inducing a net momentum in the fluid~\citep{DongEtAl2006}.

Such bifurcated wakes with jets in the time-averaged velocity field are expected for 3D wakes of low aspect ratio flapping objects~\citep{DongEtAl2006,BuchholzSmits2006,BuchholzSmits2008} and fish~\citep{TytellEtAl2010}. However, the evolution of the wake structure and vortex properties with respect to the Strouhal number, and their detailed connections to jet velocities and thrust production, have never been studied in the literature.

\section{Linking thrust production and vortex velocities}\label{sec:characteristics}

In this Section, we first focus on the vortex velocities, by relating them to the typical jet velocities and to thrust generation. 

\subsection{Measurement of vortex properties}

\begin{figure}
 \centering{\includegraphics[width=1.0\textwidth]{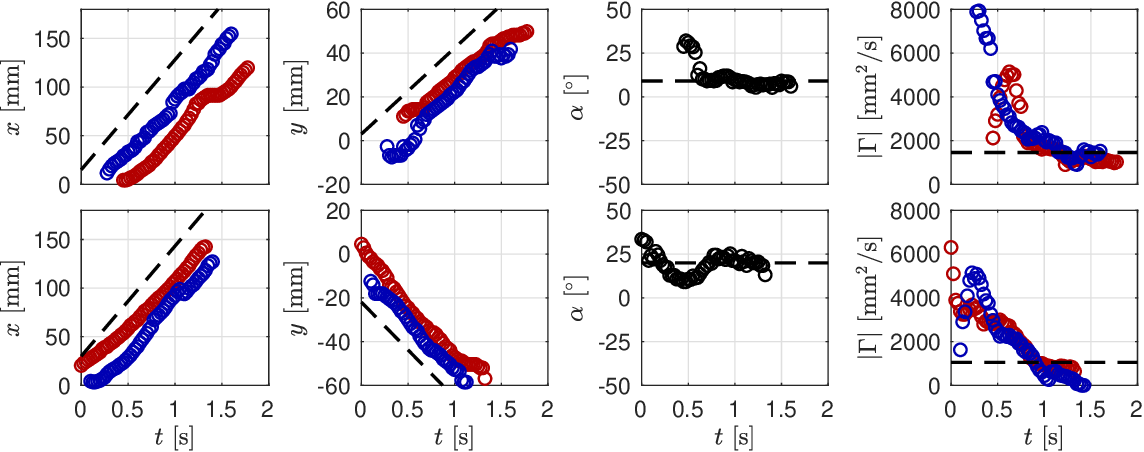}}
  \caption{Example of vortex characteristics measured as a function of time, for the 2P wake shown in Fig.~\ref{fig:4_wakes}. The top row represents the data obtained for the vortex pair in the top-half of the image, while the bottom one represents the data for the vortex pair in the bottom-half. Red points correspond to counter-clockwise vortices and blue points to clockwise ones, as the colorcode used in Fig.~\ref{fig:4_wakes}. The two first columns show the vortex position in the $(x,y)$ plane, from where the vortex velocities are extracted using a linear fit. The fitted slopes are shown with dashed lines. The two other columns represent the time evolution of the angle~$\alpha$ and of the absolute value of the circulation~$\Gamma$ measured for each vortex. The averaged quantities are shown using dashed horizontal lines.}
\label{fig:vortex_meas}
\end{figure}

Vortex characteristics are measured by following the different vortical structures in the time-dependent phase-averaged vorticity field, using a method proposed by~\citet{Godoy-DianaEtAl2009}. For each experiment, the positions of the main vortical structures in the wake are tracked, using a search of local extrema in the vorticity field. Examples of vortex detection are shown in Fig.~\ref{fig:4_wakes} for different wake types. Once the vortical structures are identified, we extracted several relevant physical quantities, such as their positions in the $(x,y)$ plane and their circulation~$\Gamma$. For pairs, we also measured the orientation angle~$\alpha$ of the vortex pairs with respect to the transverse $y$-direction and the distance~$\xi$ between the two vortices of the pair. These quantities, defined in Fig.~\ref{fig:4_wakes}(b), are then followed in time. An example of the time evolution of such quantities is illustrated in Fig.~\ref{fig:vortex_meas}, for the 2P wake shown in Fig.~\ref{fig:4_wakes}.

The vortex mean velocities in the streamwise $x$-direction, $U_\textrm{vortex}$, and in the transverse $y$-direction, $V_\textrm{vortex}$, are then obtained by a linear fit of the time evolution of the vortex positions, as illustrated by the dashed lines in the two first columns in Fig.~\ref{fig:vortex_meas}. Note that the fitted values obtained for each vortical structure are then averaged to obtain a mean value and an errorbar, for each experiment. The time evolution of the angle~$\alpha$, illustrated in the third column in Fig.~\ref{fig:vortex_meas}, is also derived from the vortex positions in the $(x,y)$ plane. A mean value and an errorbar are then obtained for each experiment by combining the two panels and performing a conditional averaging in the intermediate region where the vortices are neither too close nor too far from the fish tail, typically for $x$ between $50$ and $125$~mm.

The circulation~$\Gamma$ has been calculated by integrating the vorticity field over the surface of each vortex. The typical size~$a$ of the vortex, and therefore its surface, has been determined by a Gaussian fit 
of the vorticity field in the $x$ and $y$-directions, around the center of the vortex~\citep{Godoy-DianaEtAl2009}. The time evolution of the circulation~$\Gamma$ for a 2P wake is shown in the forth column in Fig.~\ref{fig:vortex_meas}. First, since the two vortices of each pair form a single vorticity ring in 3D, one expect the absolute values of the circulation to be equal, which is the case. Second, one can clearly see a strong decay of the circulation as the vortices get away from the tail, particularly after they have been emitted by the tail. This is clearly a 3D effect, since close to the tail, two vorticity rings are very close to each other: the one that just get ejected by the tail and the one that is currently under formation~\citep{BuchholzSmits2008}. The decay of the circulation far from the tail is more moderate. As for the angle~$\alpha$, a mean value and an errorbar are obtained for the circulation $\Gamma$ thanks to a conditional averaging in the intermediate region where $x$ is between $50$ and $125$~mm. 
Note that due to the moderate decay of the circulation in this region, the errorbars obtained may be rather large.

\subsection{Comparison between vortex and jet velocities}\label{sec:vort_jet}

\begin{figure}
 \centering{\includegraphics[width=0.75\textwidth]{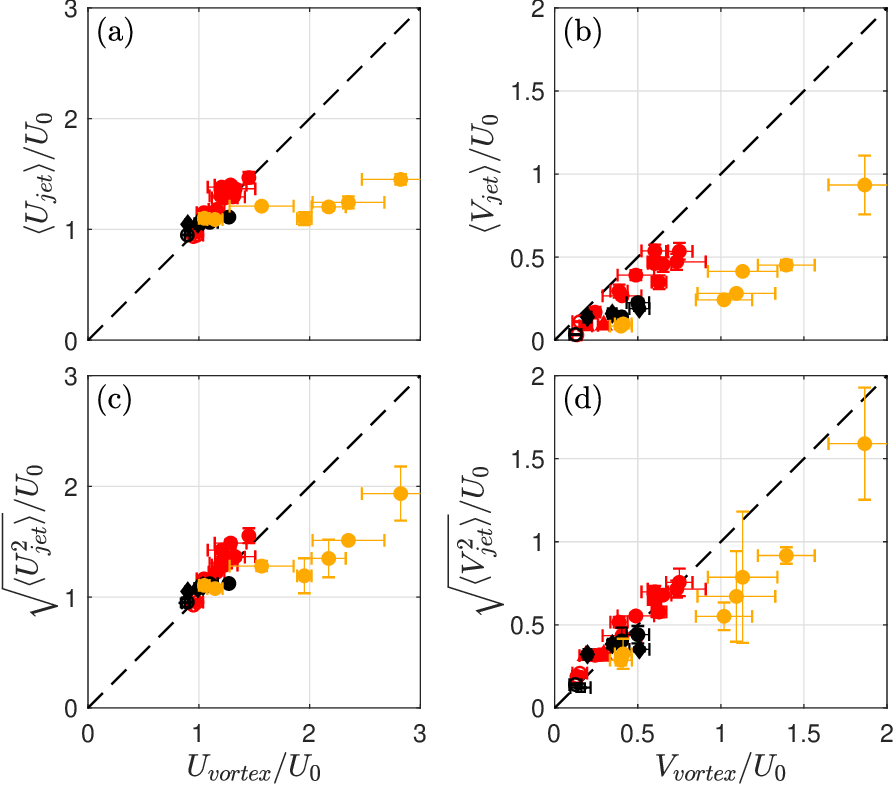}}
  \caption{Jet vs vortex velocities. In $y$-coordinates, (a) and (b) panels show the mean jet velocities in the streamwise and transverse directions, respectively, while (c) and (d) panels show the root-mean-square jet velocities. All quantities are normalised by the free-stream speed~$U_0$. The dashed line in all panels shows the unit line and is used to guide the eyes.}
\label{fig:vortex_VS_jet}
\end{figure}

The bottom panels of Fig.~\ref{fig:4_wakes} nicely shows that the vortex trajectories completely surround the jets in the time-averaged velocity field. Indeed, the travel of the vortices in the wake is responsible for the creation of these jets in the mean velocity fields, since the vortices regularly pass in the wake and locally induce a different speed than the free-stream speed in a different direction. In order to fully understand the relationship between the jets and the vortex structures, the velocities of the two objects are compared quantitatively in Fig.~\ref{fig:vortex_VS_jet}. The jet velocities are measured along $y$ profiles at given $x$ coordinates, as shown in the middle column in Fig.~\ref{fig:4_wakes}. In these profiles, the jet velocity is averaged in the region delimited by the trajectories of the vortices. 
Indeed, as one can see in Fig.~\ref{fig:4_wakes}, the vortex trajectories almost correspond spatially to the region where the time-averaged streamwise velocity field is equal to the free-stream speed (in white with the colormap used). The location of these profiles is varied in an intermediate region, where $x \in [50,125]$~mm. This thus leads to a mean jet velocity and an errorbar for each experiment. For 2S+2P and 4P wakes, we have measured the velocity of the jets formed by the main vortex pairs only. 
Figures~\ref{fig:vortex_VS_jet}(a) and~(b) show the comparison between the time-averaged jet velocities and the vortex velocities, in both streamwise $x$- and transverse $y$-direction. For the streamwise velocity, the agreement between the two is good, except for the wakes with the highest normalised vortex velocities, corresponding to the experiments with St$_A>1$, where the vortex velocity is significantly larger than the jet velocity. For the transverse velocity, the trend is relatively good but a slight offset is visible, since $V_\textrm{vortex}$ is systematically higher than $\langle V_\textrm{jet} \rangle$. Again, the largest difference is for the wakes with St$_A>1$. A careful examination of these 2P wakes at high Strouhal number shows that the vortex pairs are still identifiable but weaker in vorticity. This may be due to the fact that the wake acquires a more complex 3D structure, as also observed by~\citet{BuchholzSmits2006,BuchholzSmits2008}. Thus, the pairs are less efficient to generate jets in the horizontal mid-plane, but still travel at a high speed. Despite this phenomenon at high Strouhal number, the agreement between vortex and jet velocities is rather good for wakes at Strouhal number smaller than 1.

However, it is necessary to pay attention to the way the jet velocities are averaged in time. Indeed, since vortical structures are emitted periodically by the tail, the velocity field owns both a mean and a fluctuating components. These two components have to be taken into account for the time average, since they both influence the average pressure field~\citep{BohlKoochesfahani2009}. Therefore, we now compare the root-mean-square jet velocities~(${\langle U_\textrm{jet}^2 \rangle^{1/2}}$ and ${\langle V_\textrm{jet}^2 \rangle^{1/2}}$) with the vortex velocities. Figures~\ref{fig:vortex_VS_jet}(c) and (d) shows that these velocities are very close to the vortex velocities, except for the wakes at high Strouhal numbers. 
In addition, the differences between panels~(a) and (c) and panels~(b) and (d) in Fig.~\ref{fig:vortex_VS_jet} show that the fluctuating component of the velocity field is mainly negligible in the streamwise direction, while it cannot be neglected in the $y$-direction. This can be explained by the fact that the vortex pairs are mainly oriented towards the $y$-direction, i.e. with $\alpha$ close to~$0$ (see Fig.~\ref{fig:4_wakes}), thus having a significant effect on the pulsating component in this direction. In the following, we therefore consider that
\begin{equation}
U_\textrm{vortex}=\sqrt{\langle U_\textrm{jet}^2 \rangle} \approx \langle U_\textrm{jet} \rangle~~~~\textrm{and}~~~~V_\textrm{vortex}=\sqrt{\langle V_\textrm{jet}^2 \rangle}.\label{eq:jet_vortex}
\end{equation}
Note that Eq.~\eqref{eq:jet_vortex} for the streamwise velocity implies that the jet inversion, defined as $\langle U_\textrm{jet} \rangle=U_0$, is therefore equivalent to $U_\textrm{vortex}=U_0$.

\subsection{Vortex velocities as a function of the Strouhal number}

\begin{figure}
 \centering{\includegraphics[width=0.5\textwidth]{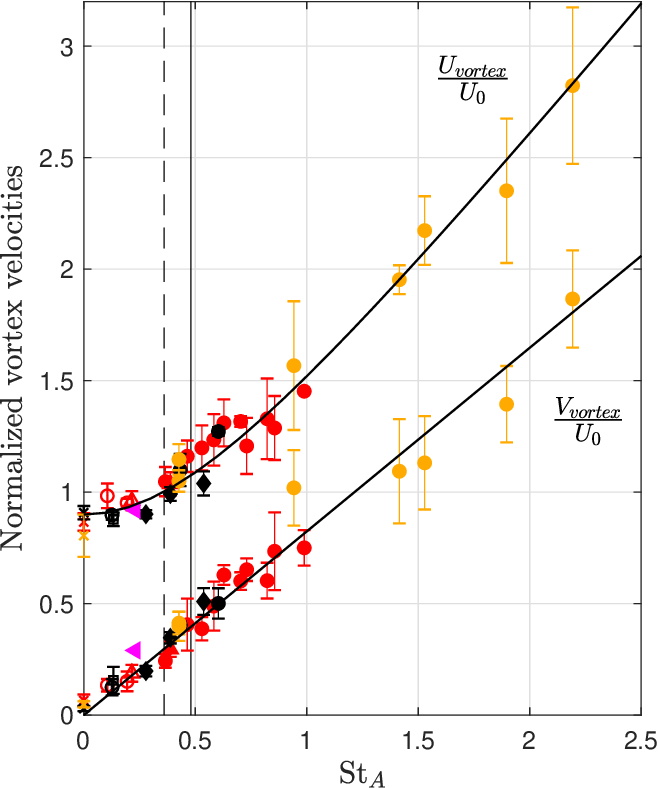}}
  \caption{Normalised streamwise ($U_\textrm{vortex}/U_0$) and transverse ($V_\textrm{vortex}/U_0$) vortex velocities as a function of the Strouhal number~St$_A$. The colors and shapes of the points are the same as the ones in Fig.~\ref{fig:param_space}. The two solid curves show the fits of the two vortex velocities, using Eqs.~\eqref{eq:fit_Vvortex} and~\eqref{eq:fit_Uvortex}. The vertical solid line at St$_A^\star=0.48$ indicates the drag-thrust transition measured with the force sensor while the vertical dashed line at St$_A^\textrm{inv}=0.36$ shows the jet inversion where $ U_\textrm{vortex} = U_0$.}
\label{fig:phase}
\end{figure}

Since jet and vortex velocities are well connected, we now focus on the evolution of the normalised vortex velocities with the Strouhal number, as shown in Fig.~\ref{fig:phase}. All points collapse well on two master curves, one for the streamwise velocity~$U_\textrm{vortex}/U_0$ (at the top) and one for the transverse velocity~$V_\textrm{vortex}/U_0$ (at the bottom). This clearly shows that the Strouhal number is the relevant control parameter for the vortex velocities. In addition, two points (one for each velocity direction) have been added from the literature (as a magenta triangle), since we were able to estimate the vortex velocities from a time series of images from~\citet{BuchholzSmits2006}. These literature points are consistent with our experimental points and lie well on the two master curves.

In Fig.~\ref{fig:phase}, the transverse vortex velocity~$V_\textrm{vortex}/U_0$ seems to be proportional to the Strouhal number. Such behavior can be easily understood since the typical dimensional scaling for $V_\textrm{vortex}$ is expected to be given by tail amplitude times tail frequency, such as $V_\textrm{vortex} \sim Af$. This leads to $V_\textrm{vortex}/U_0 \sim \textrm{St}_A$, as clearly visible in Fig.~\ref{fig:phase}. All the experimental points collapse along a line, passing through the origin. A linear fit {constrained to the origin} of these points leads to 
\begin{equation}
\frac{V_\textrm{vortex}}{U_0} = K_v\,\textrm{St}_A,\label{eq:fit_Vvortex}
\end{equation}
with $K_v \approx 0.82$, and is shown by a line {in Fig.~\ref{fig:phase}}. 

We now focus on the streamwise vortex velocity $U_\textrm{vortex}$. In Fig.~\ref{fig:phase}, $U_\textrm{vortex}/U_0$ is first almost constant and slightly below unity at very low Strouhal numbers, before increasing at high Strouhal numbers. This behavior is explained in the following model. Since we have shown that the vortex and the time-averaged jet velocities are strongly coupled, we perform a momentum balance on the mean velocity field, on a control surface surrounding the fish~\citep{Batchelor1967,Koochesfahani1989,AndersonEtAl1998,Godoy-DianaEtAl2008,BohlKoochesfahani2009}. This surface is chosen as the surface of a cuboid encompassing the fish and the wake behind it in all dimensions. Therefore, its width is chosen to be equal to the width of the imaging section in the $y$-direction, its length to be equal to the sum of the fish length~$L$ and the length of the imaging section in the streamwise $x$-direction and its thickness~$H$ of the order of the fish height in the spanwise $z$-direction. Following~\citet{Batchelor1967}, the time-averaged resultant in the $x$-direction of the different fluid forces is given by
\begin{equation}
\langle F \rangle= \langle\int(p_1 + \rho u_1^2-p_2-\rho u_2^2)\textrm{d}\Sigma-\rho\int u (\vec{u}\cdot\vec{n})\textrm{d}S\rangle,\label{eq:Batchelor1}
\end{equation}
where $u_1$, $p_1$, $u_2$ and $p_2$ are the values of the $x$-component of the velocity field or the pressure at the upstream and downstream faces respectively. $\Sigma$ is the area of these two faces, perpendicular to the streamwise direction, while $S$ is the area of the other faces of the cuboid, perpendicular to the $y$ or $z$-directions. On these latter faces, it is assumed that the flow is very close to free-stream conditions, i.e. that $\vec{u}=U_0 \vec{e}_x$. This cancels the integral on the surface $S$ in Eq.~\eqref{eq:Batchelor1}.

By applying Bernoulli's theorem~\citep{Batchelor1967,BohlKoochesfahani2009} and neglecting the velocity in the spanwise direction, Eq.~\eqref{eq:Batchelor1} becomes
\begin{equation}
\langle F \rangle=\langle\frac{1}{2}\rho\int(u_1^2-v_1^2 - u_2^2 + v_2^2)\textrm{d}\Sigma\rangle,\label{eq:Batchelor2}
\end{equation}
where $v_1$ and $v_2$ are the values of the $y$-component of the velocity field at the upstream and downstream faces. We now assume that the velocity passing through the upstream face is~$U_0$, i.e. that $\langle u_1^2 \rangle=U_0^2$ and $\langle v_1^2 \rangle=0$. At the downstream face, we consider that the velocity is also equal to the free-stream speed, i.e. that $\langle u_2^2 \rangle=U_0$ and $\langle v_2^2 \rangle=0$, except in the jets forming the V-shaped wake, where  $\langle u_2^2\rangle=\langle U_\textrm{jet}^2\rangle$ and $\langle v_2^2\rangle=\langle V_\textrm{jet}^2 \rangle$. The temporal average to consider for the mean field is therefore the one that takes into account both mean and fluctuating components of the velocity field~\citep{BohlKoochesfahani2009}, as demonstrated in section~\ref{sec:vort_jet}. Note that this simplified mean velocity field does not conserve mass in the mid-span plane, while mass conservation is classically used in the literature in such momentum balance~\citep{Batchelor1967, Godoy-DianaEtAl2008,AndersonEtAl1998}. However, as the wake structure is fully tridimensional in our case, it is not an issue to not conserve mass in a plane. Mass is of course conserved in the volume considered, but the velocity fields in the top-span and bottom-span planes, at the upper and lower limits of the fish tail in the spanwise direction, are probably very different from the one considered here as a simplification. These assumptions lead to 
\begin{equation}
\langle F \rangle=\frac{1}{2}\rho(U_0^2-\langle U_\textrm{jet}^2\rangle+\langle V_\textrm{jet}^2\rangle)2\ell H,\label{eq:Batchelor3}
\end{equation}
with $\ell$ the width of the jets along the $y$-direction, defined in Fig.~\ref{fig:4_wakes}(b). Since the vortex velocities are equal to the root-mean-square jet velocities (Eq.~\eqref{eq:jet_vortex}), except for the wakes at high Strouhal number, and with $V_\textrm{vortex}/U_0 = K_v \textrm{St}_A$ (Eq.~\eqref{eq:fit_Vvortex}), one can therefore write that 
\begin{equation}
\langle F \rangle=\rho \ell H U_0^2\left(1- \left(\frac{U_\textrm{vortex}}{U_0}\right)^2+ K_v^2 \textrm{St}_A^2\right).\label{eq:Batchelor4}
\end{equation}

Since the fish position is fixed in the laboratory frame, the force~$\langle F \rangle$ is the one measured by the force sensor, also given in Eq.~\eqref{eq:force_sensor}. Combining Eqs.~\eqref{eq:force_sensor} and~\eqref{eq:Batchelor4}
leads to
\begin{equation}
\frac{U_\textrm{vortex}}{U_0} =\sqrt{1-\frac{C_D L^2}{\ell H}+\left(K_v^2+\frac{C_D L^2}{\ell H (\textrm{St}_A^\star)^2}\right)\textrm{St}_A^2}.\label{eq:U_phase}
\end{equation}
Here, all quantities are known except the ratio~$C_D L^2/(\ell H)$. One therefore fits the experimental points for $U_\textrm{vortex}/U_0$ in Fig.~\ref{fig:phase} with Eq.~\eqref{eq:U_phase} using a single fit parameter. This leads to $C_D L^2/(\ell H) \approx 0.19$ and to
\begin{equation}
\frac{U_\textrm{vortex}}{U_0} = \sqrt{K_0^2+K_u^2\,\textrm{St}_A^2},\label{eq:fit_Uvortex}
\end{equation}
with $K_0^2 =1- C_D L^2/(\ell H)\approx 0.81$ and $K_u^2 = K_v^2+C_D L^2/(\ell H (\textrm{St}_A^\star)^2)\approx1.50$. 
This fit describes well the trend observed experimentally, as shown by the curve {in Fig.~\ref{fig:phase}}. In particular, it nicely captures that $U_\textrm{vortex}/U_0$ is almost constant at low St$_A$, close to the value where the fish is not moving at all. With a measured drag coefficient $C_D L^2=1.1 \times 10^{-3}$~m$^2$~\citep{GibouinEtAl2018}, a width of the jet of about $\ell \approx 40$~mm and a typical scale of the wake in the spanwise direction of the order of the size of the fish tail in this direction $H=60$~mm, one expects $C_D L^2/(\ell H)\approx0.45$. This represents the good order of magnitude, 
with a slight difference with the fit value associated with the fact that the wake is fully three-dimensional, with strong edge effects far from the mid-span plane. However, it is remarkable that the shape of the vortex velocity curve as a function of the Strouhal number is recovered, by measuring the velocity field in the mid-span plane only.

Regarding the jet inversion, which is here characterised by $U_\textrm{vortex}=U_0$ (see Eq.~\eqref{eq:jet_vortex}), Eq.~\eqref{eq:U_phase} shows that it occurs for a Strouhal number given by
\begin{equation}
\textrm{St}_A^\textrm{inv} = \frac{\textrm{St}_A^\star}{\sqrt{1+\frac{K_v^2 \ell H}{C_D L^2}(\textrm{St}_A^\star)^2}}.\label{eq:inversion}
\end{equation}
For $K_v > 0$, Eq.~\eqref{eq:inversion} shows that the critical Strouhal number~St$_A^\star$ for the drag-thrust transition is always larger than the Strouhal number~St$_A^\textrm{inv}$ where the jet inversion occurs. This is consistent with what has been observed in the literature of 2D wakes~\citep{Godoy-DianaEtAl2008,Godoy-DianaEtAl2009,BohlKoochesfahani2009,DasEtAl2016,AndersenEtAl2017} and here highlights the importance of the vortex velocities in the transverse $y$-direction. Using the values of $K_v\approx 0.82$ and $C_D L^2/{\ell H}\approx 0.19$ obtained from Eqs.~\eqref{eq:fit_Vvortex} and~\eqref{eq:fit_Uvortex}, and St$_A^\star \approx 0.48$, we obtain St$_A^\textrm{inv}\approx 0.36$. This value is shown as a dashed line in Fig.~\ref{fig:phase}, while the drag-thrust transition based on the force sensor at St$_A^\star \approx 0.48$ is represented by a solid line, and is indeed slightly higher. At this latter Strouhal number, Eq.~\eqref{eq:fit_Uvortex} gives $U_\textrm{vortex}^\star/U_0 \approx 1.1$, as observed in Fig.~\ref{fig:phase}. 

\section{Linking vortex properties and wake structure}\label{sec:V-shaped}
 
In this Section, we now focus on the wake structure. We first show how the wake structure is related to the vortex velocities using a simple geometrical model, and then test this model with our measurements of the wake angle, vortex pair orientation, diameter, and vorticity as a function of the Strouhal number.
 
\subsection{Wake structure model}\label{sec:structure}

Let us consider the sketch of a 2P wake shown in the horizontal mid plane behind the fish, as shown in Fig.~\ref{fig:4_wakes}(b). Each vortex pair corresponds to the 2D trace of a 3D vortex ring, has an orientation with respect to the $y$ direction given by the angle~$\alpha$, and a circulation~$\pm\Gamma$ for each vortex. Once generated, the pair has a self-propelled speed, noted~$U_\textrm{dipole}$, that depends on its circulation~$\Gamma$ and on its geometry~\citep{GuyonEtAl2001}.
In our case, the fish position is fixed in the laboratory frame but a free-stream flow speed~$U_0$ is imposed. Therefore, the vortex pair velocity in the laboratory frame results from the vector combination of the free-stream velocity and the self-induced velocity of the vortex pair.
Even though the formation of the vortical structures in the vicinity of the tail is a very complex phenomenon~\citep{DongEtAl2006,BuchholzSmits2006,BuchholzSmits2008,LiDong2016}, we assume for simplicity that a vortex pair is shed at time~$t$ at the end of the fish tail, in position~$A$. At time $t+ \Delta t$, this pair has moved to position~$B$, driven by the free-stream speed~$U_0$ in the streamwise direction and by its self-induced speed~$U_\textrm{dipole}$ in an oblique direction at an angle~$\alpha$.
Therefore, its trajectory~$AB$ can be decomposed as $AC = U_0 \Delta t$ and $CB = U_\textrm{dipole} \Delta t$. By repeating this operation for different~$\Delta t$, all vortex pairs align on two symmetrical straight lines at an angle~$\theta$ with respect to the axis of the wake~\citep{CouderBasdevant1986,DongEtAl2006}. The angle~$\theta$ is given by
\begin{equation}
\tan \theta = \frac{V_\textrm{vortex}}{U_\textrm{vortex}},\label{eq:theta_vortex}
\end{equation}
where the vortex velocities in the laboratory frame can be expressed through the combination of advection and self-induced velocities:
\begin{eqnarray}
U_\textrm{vortex}&=&U_0+U_\textrm{dipole} \sin \alpha,\label{eq:Uvortex}\\
V_\textrm{vortex}&=&U_\textrm{dipole} \cos \alpha,\label{eq:Vvortex}
\end{eqnarray}
respectively in $x$ and $y$-directions.

Equations~\eqref{eq:theta_vortex} to~\eqref{eq:Vvortex} show that the wake angle and vortex properties are indeed intimately all interconnected. In the following of the Section, we demonstrate that this simplified framework is valid for the wakes observed in this work and we explain how the different vortex characteristics contribute to shape the wake. Note that this description has been performed for a 2P wake but most of the considerations and definitions can be at least extended to the 2S+2P and the 4P wakes observed and described in this work. Indeed, for these two types of wakes, one can see that there is a main pair of vortices, propagating along the oblique thick lines, and an extra vortex (for 2S+2P) or an extra pair (for 4P). As illustrated in Fig.~\ref{fig:4_wakes}, one can clearly see that these extra vortical structures travel together at a similar speed and same oblique angle {$\theta$} than the main pair, within the observation window used in this study. By applying the definitions mentioned above to the main pair, we can generalise this view to 2S+2P and 4P wakes. For the 2S wake, this view cannot be applied since it is composed of horseshoe vortices, that interact with each other with their vorticity in the streamwise direction~\citep{BuchholzSmits2006}, which is not the case once vortex rings are formed in the 2P case~\citep{BuchholzSmits2008}.

\subsection{Wake angle}

\begin{figure}
  \centering{\includegraphics[width=0.8\textwidth]{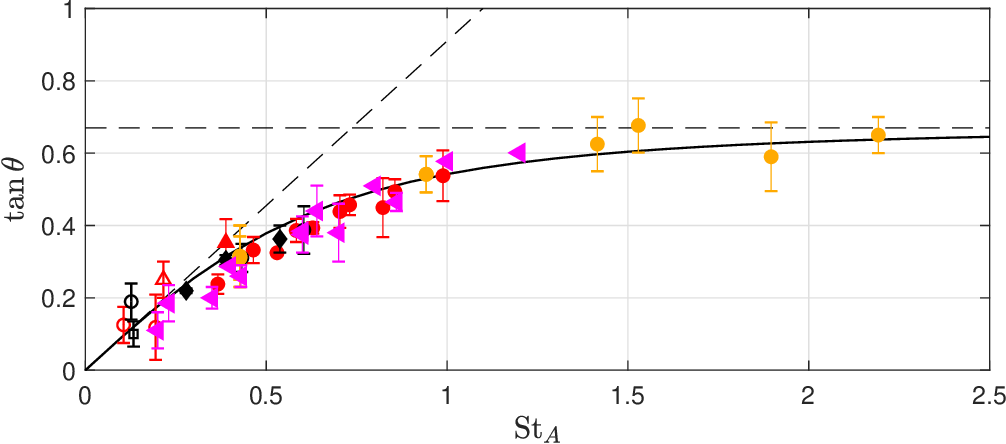}}
  \caption{Tangent of the wake angle~$\theta$ as a function of the Strouhal number~St$_A$. The colors and shapes of the points are the same as the ones in Fig.~\ref{fig:param_space}, while the magenta triangles correspond to literature data. The dashed lines indicate $\tan \theta = K_v/K_0\,\textrm{St}_A\approx 0.91\,\textrm{St}_A$ and $\tan \theta = K_v/K_u\approx 0.67$, while the curve is defined by Eq.~\eqref{eq:theta_Strouhal}.}
\label{fig:angle_wake}
\end{figure}

As shown in Eq.~\eqref{eq:theta_vortex}, the wake angle~$\theta$ is defined as the angle formed between the vortex trajectories and the streamwise $x$-direction. Combining the two fits~\eqref{eq:fit_Vvortex} and~\eqref{eq:fit_Uvortex} obtained for $U_\textrm{vortex}/U_0$ and $V_\textrm{vortex}/U_0$ as a function of St$_A$ into Eq.~\eqref{eq:theta_vortex} for the angle~$\theta$, one obtains
\begin{equation}
\tan \theta = \frac{K_v \textrm{St}_A}{\sqrt{K_0^2+K_u^2\,\textrm{St}_A^2}}.\label{eq:theta_Strouhal}
\end{equation}
Since the vortex are responsible for the formation of the jets within the wake (see Fig.~\ref{fig:4_wakes}), the jets also exhibit the same angle~$\theta$ with respect to the streamwise direction. The wake angle~$\theta$ is therefore directly measured using the time-averaged velocity field from 2D PIV data. For each experiment, the angle of each jet behind the fish is measured, from both sides of the jet. The angles obtained for the different jets are then averaged and used to estimate error bars. An other method to measure the wake angle based on vortex trajectories led to very similar results, but with slightly larger errorbars. The tangent of~$\theta$ is plotted in Fig.~\ref{fig:angle_wake} as a function of the Strouhal number. While three different physical parameters have been varied in large ranges and four different wake patterns have been observed, all data collapse well on a master curve, showing again that the Strouhal number is the relevant dimensionless parameter that controls the wake properties. One nicely recovers the master curve by plotting $\tan \theta$ defined in Eq.~\eqref{eq:theta_Strouhal} in Fig.~\ref{fig:angle_wake}. The master curve exhibits two limit regimes, shown by the two dashed lines. At low Strouhal number, {$\tan \theta$ is proportional to $\textrm{St}_A$} since $U_\textrm{vortex}/U_0$ is almost constant and $V_\textrm{vortex}/U_0$ is linear with the Strouhal number. At large Strouhal number, the angle saturates to an asymptotic value, as both normalised velocities are linear with the Strouhal number. Using Eq.~\eqref{eq:theta_Strouhal}, the proportionality and saturation constants are equal to $K_v/K_0$ and $K_v/K_u$, respectively, estimated as~$0.91$ and~$0.67$ from the fits~\eqref{eq:fit_Vvortex} and~\eqref{eq:fit_Uvortex}. Note that \citet{BuchholzSmits2006} have already noticed that, at low Strouhal number, the angle of the wake increases proportionally with the Strouhal number.

In addition to our experimental measurements, several points extracted from the literature (when a 3D wake is visible and can be easily measured) have been added. These points are plotted as magenta triangles (pointing left) in Fig.~\ref{fig:angle_wake} and overlap well with the master curve, independently of their configuration: numerical simulations of fish~\citep{BorazjaniSotiropoulos2008} or moving panels~\citep{BlondeauxEtAl2005,DongEtAl2006,LiDong2016}, experiments with pitching panels~\citep{BuchholzSmits2006} or with an other robotic fish~\citep{EppsEtAl2009}. This seems to show that this master curve is universal, and is mainly independent of other parameters, such as Reynolds number, exact shape, aspect ratio or swimming mode of the moving objects.

\subsection{Orientation of the vortex rings}

We now focus on the orientation of the vortex pairs with respect to the $y$-direction. In this section, we restrict ourselves to the main vortex pair of the 2P, 2S+2P and 4P wakes, since it is possible to define and measure the proper orientation of such a vortex pair in these wakes. 

\begin{figure}
  \centering{\includegraphics[width=1\textwidth]{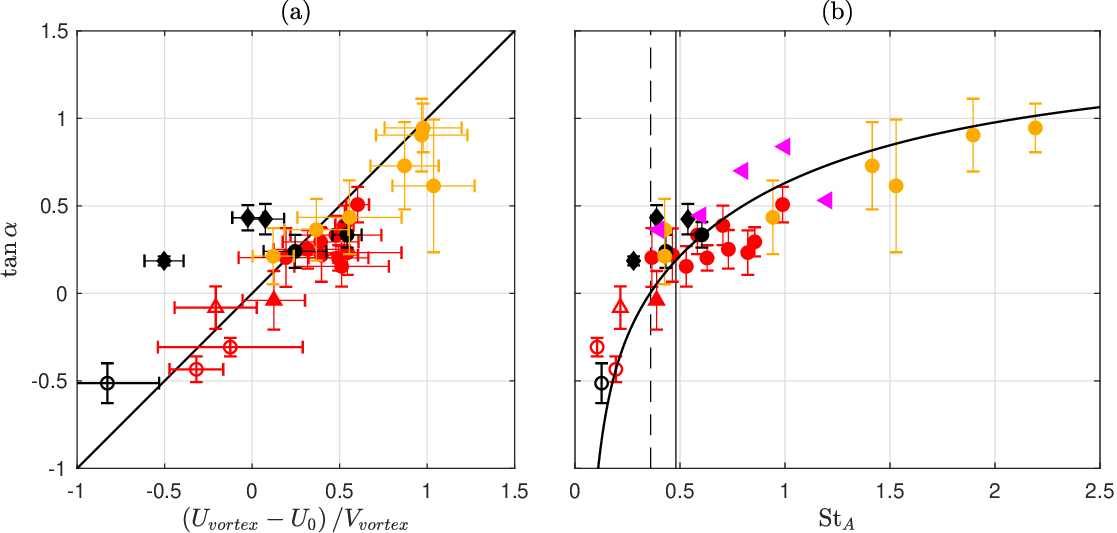}}
  \caption{Tangent of the angle $\alpha$ as a function of (a) the combination of $U_\textrm{vortex}/U_0$ and $V_\textrm{vortex}/U_0$ given in Eq.~\eqref{eq:alpha_vortex} and (b) the Strouhal number St$_A$. Only the experiments exhibiting a 2P, 2S+2P or a 4P wake are plotted. The lines show the prediction of Eq.~\eqref{eq:alpha_vortex} (panel (a)) and Eq.~\eqref{eq:alpha_Strouhal} (panel (b)). Magenta triangles in panel (b) correspond to the points given by~\citet{LiDong2016}. In panel (b), the vertical solid line indicates the drag-thrust trnasition at St$_A^\star=0.48$ while the vertical dashed line at St$_A^\textrm{inv}=0.36$ shows the vortex inversion where $ U_\textrm{vortex} = U_0$.}
\label{fig:alpha}
\end{figure}

Using Eqs.~\eqref{eq:Uvortex} and~\eqref{eq:Vvortex}, the tangent of the angle~$\alpha$ can easily be obtained as
\begin{equation}
\tan \alpha = \frac{U_\textrm{vortex}-U_0}{V_\textrm{vortex}}.\label{eq:alpha_vortex}
\end{equation}
This expression is tested in Fig.~\ref{fig:alpha}(a). It shows that Eq.~\eqref{eq:alpha_vortex} is well verified for the 2P and 4P wakes. The three points (black diamonds) corresponding to a 2S+2P wake appear as slightly shifted from the remainder of the points. On each oblique direction of this wake, there is a main vortex pair and a single vortex, that travel together at the same speed. They are probably the 2D footprint of a vortex ring and a horseshoe vortex. The horseshoe vortex interacts directly with the vortex ring due to the short distance between them, but may also interact with other structures located at a larger distance through its streamwise leg. This may influence the orientation of the vortex ring and affect its trajectory. Note that the secondary vortex ring in the 4P wake does not seem to produce such difference, maybe because it is a vortex ring and thus does not own a streamwise leg, contrary to horseshoe vortices.

With the fit expressions~\eqref{eq:fit_Vvortex} and~\eqref{eq:fit_Uvortex} for the normalised vortex velocities as a function of the Strouhal number, $\tan \alpha$ can easily be expressed as a function of the Strouhal number, through Eq.~\eqref{eq:alpha_vortex}:
\begin{equation}
\tan \alpha = \frac{\sqrt{K_0^2+K_u^2\,\textrm{St}_A^2}-1}{K_v \textrm{St}_A}.\label{eq:alpha_Strouhal}
\end{equation}
Figure~\ref{fig:alpha}(b) shows the comparison between our experimental measurements of~$\tan \alpha$, the data from~\citet{LiDong2016} (magenta triangles) and Eq.~\eqref{eq:alpha_Strouhal}, as a function of the Strouhal number. Again, all data collapse well on the expected behavior. In particular, it nicely shows that the angle $\alpha$ of the wake increases from negative to positive value with increasing Strouhal number. According to Eq.~\eqref{eq:alpha_vortex}, the angle~$\alpha$ is equal to~$0$ when $U_\textrm{vortex}=U_0$, i.e. when the Strouhal number reaches the inversion Strouhal number~St$_A^\textrm{inv}\approx 0.36$, defined in Eq.~\eqref{eq:inversion}. This corresponds to the vortex or jet inversion, where the vortex pair orientation changes from negative to positive or where the mean streamwise component of the jet velocity is equal to the free-stream speed. As shown by the vertical dashed dotted lines at St$_A^\textrm{inv}= 0.36$ in Figs.~\ref{fig:phase} and~\ref{fig:alpha}(b), $U_\textrm{vortex}=U_0$ and $\alpha=0$ occur at the same Strouhal number. Since the inversion Strouhal number is only slightly smaller than the drag-thrust transition Strouhal number, the orientation of the vortex pairs is a good indicator of the propulsion regime (drag or thrust), as already noticed in the literature for 2D wakes~\citep{AndersenEtAl2017} and 3D wakes~\citep{DongEtAl2006}. However, the drag-thrust transition does not occur exactly for $\tan \alpha=0$ but for $\tan \alpha \approx 0.09$ here, i.e. $\alpha\approx5^\circ$, using Eq.~\eqref{eq:alpha_Strouhal}.
\subsection{Circulation and diameter of the vortex rings}

\begin{figure}
  \centering{\includegraphics[width=1\textwidth]{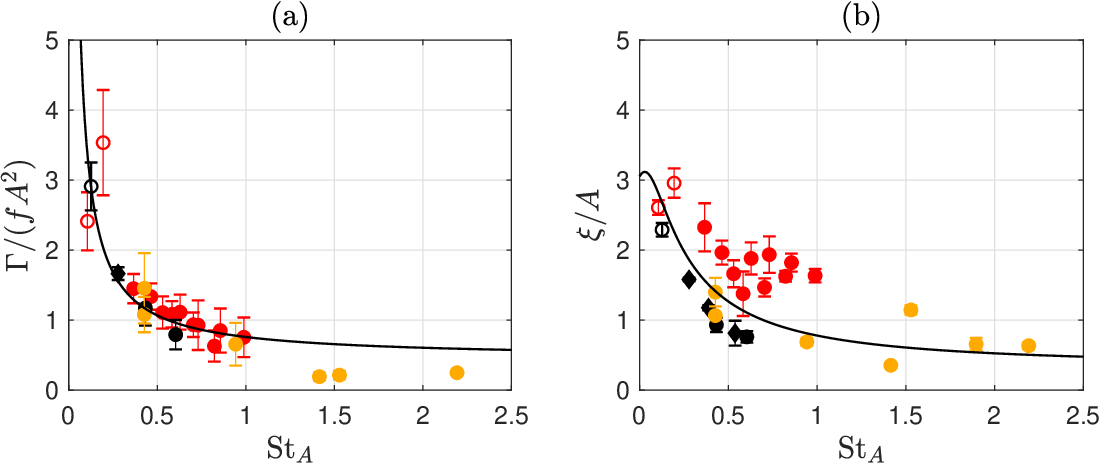}}
  \caption{Dimensionless circulation~$\Gamma/(fA^2)$ (a) and pair diameter~$\xi/A$ (b) as a function of the Strouhal number~St$_A$, for 2P wakes. The solid curve in panel (a) shows the fit of the data by Eq.~\eqref{eq:gamma}, while the one in panel (b) represents Eq.~\eqref{eq:xi_Strouhal}.}
  \label{fig:gamma}
\end{figure}

We now discuss the measurements of the vortex ring circulation~$\Gamma$. We limit ourselves to the 2P wakes observed in this study, since having more vortices in the wake may significantly change the circulation magnitude of the main vortex pair. \citet{BuchholzEtAl2011} have proposed a scaling for the circulation in the case of a pitching panel with low aspect ratio, i.e. in the presence of a 3D wake. This kinematic scaling originates from the production of vorticity by the motion of the panel, having a transverse speed~$Af$ on a typical time~$1/f$. They have found experimental evidences that the circulation~$\Gamma$ mainly scales as $f A^2$, with extra corrections depending on the ratio between the amplitude~$A$ and the height of the panel in the spanwise direction. 
In their experiments, they have varied these two parameters and the free-stream speed~$U_0$ but the oscillation frequency~$f$ has been kept constant. In our case, the circulation have been measured for all experiments, where the three main parameters~$(A,f, U_0)$ have been varied significantly while keeping the geometry of the tail constant. This has allowed us to test how the circulation varies with these three parameters with respect to several hypotheses. Figure~\ref{fig:gamma}(a) shows the vortex ring circulation~$\Gamma$ normalised by the scaling $f A^2$ as a function of the Strouhal number~St$_A$, for the 2P wakes observed in this study. One clearly observes that the normalised circulation decreases with the Strouhal number, while it is expected to be constant according to the scaling of \citet{BuchholzEtAl2011}.

In our case, it is therefore necessary to consider another scaling for the circulation. Indeed, even if the motion of the tail clearly contributes to the creation of vorticity, one can also consider that vorticity can be created when the tail is at a fixed angle in a streamwise flow. Even though the angle of the tail is in average equal to zero, the tail spends a significant amount of time close to its extremum position during the oscillation. One can therefore expect to have two contributions for the scaling of the circulation~$\Gamma$: a kinematic scaling~$fA^2$~\citep{BuchholzEtAl2011}, independent of the streamwise speed~$U_0$, and a stationary scaling~$U_0 A$, independent of the frequency motion~$f$. When normalising the circulation by the kinematic scaling as in Fig.~\ref{fig:gamma}(a), this leads to
\begin{equation}
\frac{\Gamma}{fA^2}=\gamma_d + \frac{\gamma_s}{\textrm{St}_A},\label{eq:gamma}
\end{equation}
where $\gamma_d$ and $\gamma_s$ are two dimensionless coefficients, reflecting the contribution of the kinematic and stationary scalings respectively. The data represented in Fig.~\ref{fig:gamma}(a) can be nicely fitted by Eq.~\eqref{eq:gamma}, leading to $\gamma_d=0.21$ and $\gamma_s=0.45$. These two coefficients are of the same order of magnitude but, for a Strouhal number smaller than~$1$, the dominant contribution comes from the stationary scaling. In the experiments of \citet{BuchholzEtAl2011}, the stationary contribution appears negligible. This may be due to the fact that the undulating motion of the robotic fish is more complex than a simple pitching motion used by \citet{BuchholzEtAl2011}. 

The circulation~$\Gamma$ therefore seems to scale depending on a stationary configuration, while its magnitude clearly affects the wake structure and the thrust production, that are related to the instationarity of the wake. Indeed, the circulation~$\Gamma$ is connected to the vortex velocities through the pair self-advecting speed~$U_\textrm{dipole}$ and the pair diameter~$\xi$. A comprehensive calculation~\citep{GuyonEtAl2001}, assuming that the vortex owns a cylindrical core of size~$a$ in uniform rotation, gives
\begin{equation}
U_\textrm{dipole}=\frac{\Gamma}{2\pi\xi}\left(\ln \frac{4\xi}{a}-\frac{1}{2}\right),\label{eq:U_dipole}
\end{equation}
and is consistent with the typical simplified scaling $U_\textrm{dipole}\sim \Gamma/\xi$.

The distance~$\xi$, normalised by the tail amplitude~$A$, is plotted in Fig.~\ref{fig:gamma}(b) as a function of the Strouhal number, {showing a decreasing trend with increasing values of this parameter.} 
To understand this trend, one can express the self-induced speed from Eqs.~\eqref{eq:Uvortex} and \eqref{eq:Vvortex}:
\begin{equation}
U_\textrm{dipole}=\left(\left(U_\textrm{vortex}-U_0\right)^2+V_\textrm{vortex}^2\right)^{1/2},\label{eq:U_dipole2}
\end{equation}
and combine the fits~\eqref{eq:fit_Vvortex},~\eqref{eq:fit_Uvortex} and~\eqref{eq:gamma} of the normalised vortex velocities and circulation as a function of the Strouhal number together in the simplified scaling of Eq.~\eqref{eq:U_dipole}:
\begin{equation}
\frac{\xi}{A} \approx \frac{\Gamma}{A U_\textrm{dipole}}=\frac{\Gamma\textrm{St}_A}{fA^2}\frac{U_0}{ U_\textrm{dipole}}=\frac{\gamma_d \textrm{St}_A+ \gamma_s}{\left(\left(\sqrt{K_0^2+K_u^2\,\textrm{St}_A^2}-1\right)^2+K_v^2 \textrm{St}_A^2\right)^{1/2}}.\label{eq:xi_Strouhal}
\end{equation}
The prediction of Eq.~\eqref{eq:xi_Strouhal}, shown as a solid curve in Fig.~\ref{fig:gamma}(b), captures rather well the experimental trend. At small Strouhal number, i.e. St$_A<0.5$, Eq.~\eqref{eq:xi_Strouhal} leads to $\xi/A \approx \gamma_s /(K_v\textrm{St}_A)$, consistent with the decrease of $\xi/A$ for small Strouhal in Fig.~\ref{fig:gamma}(b). At large Strouhal number, i.e. St$_A>1$, Eq.~\eqref{eq:xi_Strouhal} leads to the convergence of $\xi/A$ towards a constant~$\gamma_s/\sqrt{K_u^2+K_v^2}\approx 0.3$, also consistent with Fig.~\ref{fig:gamma}(b) at large Strouhal number. Note that Eq.~\eqref{eq:xi_Strouhal} is obtained by using the simple scaling law~$U_\textrm{dipole}=\Gamma/\xi$, instead of using the full Eq.~\eqref{eq:U_dipole}. Indeed, the factor~$\ln\left((4\xi/a)-1/2\right)/(2\pi)$ has a weak dependence on Strouhal number, and has been measured around~$0.4$ in our experiments. A factor of value~$1$ better adjusts our experimental measurements for~$\xi$, but the order of magnitude is the same and Eq.~\eqref{eq:U_dipole} assumes a vortex ring with a cylindrical core in uniform rotation, which may be not the case for the vortices in the wake.

\section{Conclusions}\label{sec:ccl}

In this paper, using extensive data processing of 2D PIV measurements in the mid-span plane behind a tethered robotic fish immersed in a free stream, we have studied the wake properties of the produced vortex streets and their connections with thrust production. We have mainly observed 2P wakes in our experiments, but we have shown that the described trends remain largely valid for other types of wakes, such as 2S, 2S+2P, 4P.

First, we have focused on the velocities of the vortical structures in the mid-span plane. We have shown that the jets formed within the wake and visible in the time-averaged velocity field are produced by the vortical structures as they periodically pass after their emission by the tail. In particular, we have experimentally shown that the root mean square of the jet velocites are equal to the vortex velocities. In addition, we have understood how these vortex velocities vary with the Strouhal number. These variations have been described by a model using a momentum balance, allowing to establish a clear correspondence between vortex velocities and thrust production. 
In particular, we have demonstrated that the jet inversion (defined as $\langle U_\textrm{jet} \rangle =U_0$) occurs for a Strouhal number slightly~St$_A^\textrm{inv}$ smaller that the critical Strouhal number~St$_A^\star$ where the drag/thrust transition takes place, thus explaining what has already been observed in the literature. Through Eq.~\eqref{eq:inversion}, it appears  that this slight shift in Strouhal number is due to the component of the vortex or jet velocity in the transverse $y$-direction. This component itself originates from the downstream pressure term in Eq.~\eqref{eq:Batchelor1} and appears in Eq.~\eqref{eq:Batchelor2} thanks to the application of the Bernoulli's theorem.

Second, we have developed a simple geometrical model for the wake structure, in connection with vortex velocities. This model considers that the vortex pair speed is composed of the free-stream speed and the pair self-advecting speed. It therefore connects the wake angle~$\theta$ and the orientation angle of the main vortex pair~$\alpha$ to the vortex velocities, and thus thrust production. The data obtained from our measurements and from the literature are in good agreement with the model, both for the wake angle and the orientation of the main vortex pair. This highlights a universal behavior for the 3D wakes behind fish or low aspect ratio flapping objects, mainly governed by the Strouhal number despite large differences in Reynolds number, shape, or swimming mode. 

Finally, we have also derived and experimentally tested a model for the circulation~$\Gamma$, related to the pair diameter~$\xi$ and the vortex velocities. The circulation has been found to be consistent with the different physical quantities and formulae obtained in this article. 

This work therefore allows us for a comprehensive understanding of how thrust production and wake structure are related to vortex properties in 3D wakes. In addition, this work can also be relevant to understand fish-fish interactions. Indeed, the structure of fish wake is a crucial parameter to consider in understanding fish schooling systems~\citep{Weihs1973} or predator-prey relations~\citep{Spedding2014}. Fish are sensitive to hydrodynamic signals through the use of their lateral line system~\citep{PartridgePitcher1980}, and they are therefore able to sample the vortical structures emitted by other fish in their neighbouring. Knowing the organization and properties of their vortex streets is therefore a key information to understand these interactions. For example, \citet{LiEtAl2019a} have shown in recent simulations that fish experience the largest drag when they plunge into the wake of the fish in front of them, since they directly swim in a jet opposing their motion. On the contrary, \citet{VermaEtAl2018} have shown that, in numerical simulations, fish can learn to use the vortical structures of the wake of their neighbour to swim more efficiently. Despite these discrepancies, the wake angle appears as a crucial quantity. Since fish mainly swim at a Strouhal number between 0.2 and 0.4, our measurements of the wake angle in Fig.~\ref{fig:angle_wake} indicates that the vortical structures produced by fish cruising at constant speed are shed with an angle between $11^\circ$ and $17^\circ$. This work thus paves the way towards a better understanding of fish-fish hydrodynamic interactions.


\backsection[Acknowledgements]{The authors thank Li Fu for his help regarding the fish actuation and measurements with the force sensor. Thomas Frisch is acknowledged for helpful discussions on vortex interactions during the preliminary stage of this work.}

\backsection[Funding]{This work was supported by the French government through the UCA Joint Excellent Dynamic Initiative (JEDI) Investment in the Future Programs, managed by the National Research Agency (ANR) [reference no. ANR-15-IDEX-0001]. The National Centre for Scientific Research (CNRS) is gratefully acknowledged for the initial support of Christophe Brouzet.}

\backsection[Declaration of interests]{The authors report no conflict of interest.}


\backsection[Author ORCIDs]{C. Brouzet, https://orcid.org/0000-0003-3131-3942; C. Raufaste, https://orcid.org/0000-0003-4328-7438; M. Argentina, https://orcid.org/0000-0002-8926-7398}




\bibliographystyle{jfm}
\bibliography{Fish_swimming}

\end{document}